\documentclass[journal]{IEEEtran}
\ifCLASSINFOpdf
  \usepackage[pdftex]{graphicx} 
\else
  \usepackage[dvips]{graphicx}
  declare the path(s) where your graphic files are
  \graphicspath{{../eps/}}
  
 \fi

\usepackage[cmex10]{amsmath}
\usepackage{amssymb}

\usepackage{caption}
\usepackage{subcaption}
\usepackage{float}
\usepackage[table]{xcolor}
\usepackage{multirow}
\usepackage{multicol}
\usepackage{cuted}
\usepackage{enumerate}
\usepackage{acro}
\usepackage{epstopdf}
\usepackage{cite}
\usepackage{xcolor}
\usepackage{gensymb}
\usepackage{siunitx}
\usepackage{adjustbox}
\usepackage{array}
\usepackage{graphics}
\usepackage{graphicx}
\newcolumntype{C}[1]{>{\centering\arraybackslash}m{#1}}

\usepackage{caption}
\acsetup{list-long-format=\capitalisewords}
\DeclareAcronym{PLS}{short= PLS, long= physical layer security}
\DeclareAcronym{RSS}{short= RSS, long= received signal strength}
\DeclareAcronym{CIR}{short= CIR, long= channel impulse response}
\DeclareAcronym{CFR}{short= CFR, long= channel frequency response}
\DeclareAcronym{OFDM}{short= OFDM, long= orthogonal frequency division multiplexing}
\DeclareAcronym{qos}{short= QoS, long= quality-of-service}
\DeclareAcronym{PAPR}{short= PAPR, long= peak to average power ratio}
\DeclareAcronym{aoa}{short= AoA, long= angle of 
arrival}
\DeclareAcronym{aod}{short= AoD, long= angle of departure}
\DeclareAcronym{FIR}{short= FIR, long= finite impulse response}

\DeclareAcronym{SISO}{short=SISO, long=single-input single-output}

\DeclareAcronym{IFFT}{
  short = IFFT,
  long  = inverse fast Fourier transform ,
  class = abbrev
}
\DeclareAcronym{FFT}{
  short = FFT,
  long  = fast Fourier transform ,
  class = abbrev
}
\DeclareAcronym{IDFT}{
  short = IDFT,
  long  = inverse discrete Fourier transform ,
  class = abbrev
}
\DeclareAcronym{CP}{
  short = CP,
  long  = cyclic prefix ,
  class = abbrev
}
\DeclareAcronym{ISI}{
  short = ISI,
  long  = inter-symbol interference ,
  class = abbrev
}
\DeclareAcronym{BER}{
  short = BER,
  long  = bit error rate ,
  class = abbrev
}
\DeclareAcronym{AWGN}{
  short = AWGN,
  long  = additive white Gaussian noise,
  class = abbrev
}
\DeclareAcronym{MMSE}{
  short = MMSE ,
  long  = minimum mean squared error ,
  class = abbrev
}

\DeclareAcronym{NMSE}{
  short = NMSE ,
  long  = normalized mean squared error ,
  class = abbrev
}

\DeclareAcronym{CSI}{
  short = CSI ,
  long  = channel state information ,
  class = abbrev
}

\DeclareAcronym{SNR}{
  short = SNR ,
  long  = signal-to-noise ratio ,
  class = abbrev
} 

\DeclareAcronym{QPSK}{
  short = QPSK ,
  long  = quadrature phase shift keying ,
  class = abbrev
}
\DeclareAcronym{BEP}{
  short = BEP ,
  long  = bit error probability ,
  class = abbrev
}
\DeclareAcronym{RV}{
  short = RV,
  long  = random variable ,
  class = abbrev
}

\DeclareAcronym{CDF}{
  short = CDF,
  long  = cumulative distribution function ,
  class = abbrev
}
\DeclareAcronym{SOP}{
  short = SOP,
  long  = secrecy outage probability ,
  class = abbrev
}
\DeclareAcronym{SINR}{
  short = SINR,
  long  = signal-to-interference-plus-noise ratio ,
  class = abbrev
}
\DeclareAcronym{DFT}{
  short = DFT,
  long  = discrete Fourier transform ,
  class = abbrev
}
\DeclareAcronym{PDF}{
  short = PDF,
  long  = probability density function ,
  class = abbrev
}
\DeclareAcronym{AN}{
  short = AN,
  long  = artificial noise,
  class = abbrev
}

\DeclareAcronym{5G}{
  short = 5G,
  long  = fifth-generation,
  class = abbrev
}
\DeclareAcronym{LS}{
  short = LS,
  long  = link-signature,
  class = abbrev
}

\bstctlcite{IEEEexample:BSTcontrol}

\begin{document}
\bstctlcite{IEEEexample:BSTcontrol}
    \title{Flexible Physical Layer Security for Joint Data and Pilots in Future Wireless Networks\thanks{This work has been submitted to the IEEE Transactions on Communications for possible publication. Copyright may be transferred without notice, after which this version may no longer be accessible.}}
   \author{Salah Eddine Zegrar, Haji M. Furqan,
      H\"{u}seyin Arslan,~\IEEEmembership{Fellow,~IEEE}

\thanks{The authors are with the Department of Electrical and Electronics Engineering, Istanbul Medipol University, Istanbul, 34810, Turkey (e-mail: salah.zegrar@std.medipol.edu.tr; hamadni@st.medipol.edu.tr;  huseyinarslan@medipol.edu.tr).}
\thanks{H. Arslan is also with Department of Electrical Engineering,
	University of South Florida, Tampa, FL, 33620, USA.}
}
\maketitle
\begin{abstract}
In this work, novel physical layer security (PLS) schemes are proposed for orthogonal frequency-division multiplexing (OFDM) to secure both data and pilots. 
The majority of previous studies focus on only securing the data without considering the security of the pilots used for channel estimation. However, the leakage of channel state information (CSI) from a legitimate node to an eavesdropper allows the latter to acquire knowledge about the channel of the legitimate nodes. To this end, we propose adaptive and flexible PLS algorithms which can 1) secure data, 2) secure pilots, and 3) jointly secure both data and pilots.
Particularly, minimum-phase all-pass channel decomposition is exploited, where the proposed algorithms use the all-pass component to provide security without harming the performance of the legitimate user.
In the analysis for data security, we evaluated the secrecy under correlated and uncorrelated eavesdropping channels via closed-form bit error rate (BER) formulas. For pilot security, we analyzed the normalized mean squared error (NMSE) performance of the estimated channel. The simulation results along with theoretical analysis demonstrate that the proposed algorithms can effectively enhance the communication secrecy of the overall system.
\end{abstract}

\section{Introduction}
\IEEEPARstart{W}{hile} an enormous amount of novel and efficient wireless technologies have been proposed in order to fulfill the demands of \ac{5G} and beyond in terms of reliability, throughput, and latency, the security has become a sensitive issue \cite{hu2015mobile}. This is due to the fact that the open and broadcast nature of wireless transmission makes the physical transmitted signal, bearing the communication data and sensitive information, vulnerable to eavesdropping \cite{6739367}. In order to overcome the security threats, upper-layer encryption-based algorithms are exploited conventionally. Such techniques, however, may not be feasible for future wireless networks because of the difficulties in terms of key management and sharing in these heterogeneous wireless networks. \Ac{PLS} has emerged as an interesting and powerful solution that can complement conventional security techniques and improve the overall security of wireless communication networks \cite{bloch2011physical}. Particularly, \ac{PLS} observes and exploits the dynamic characteristics of the signal, radio, and channel for ensuring the security of features and contents at the physical layer \cite{rivest1990handbook}. 

\subsection{PLS Literature Review}
\Ac{PLS} techniques are proposed to achieve two different goals, namely, securing the data communication and channel estimation. The first goal of \ac{PLS} algorithms is to degrade the data decoding capability of non-legitimate nodes compared to the legitimate node by exploiting different properties of the wireless channel. While the second goal is to enforce poor channel estimation at the eavesdropper/attacker, which will degrade the signal recovery capability at the attacker \cite{9336039, 4543070, 8509094}.

\subsubsection{PLS for Data}
Among the top areas in \ac{PLS}, securing \ac{OFDM} has drawn enormous attention recently since \ac{OFDM} is the most commonly employed waveform in the current and next-generation systems \cite{8093591}. In line with this vision, different security techniques have been proposed in the literature. These techniques include key generation-based approaches \cite{7120014,keybased2,keybased3}, adaptive communication-based approaches \cite{7562191}, and \ac{AN}-based techniques \cite{6516879,AN1}. 

\par Key generation-based techniques are based on the exploitation of channel reciprocity property between legitimate nodes as a common source of randomness. For example, amplitude and phase related to \ac{RSS}, \ac{CIR}, \ac{CFR}, and other feedback that can be used for key generation \cite{7120014} \cite{7393435}. These techniques are interesting in the sense that they can solve key management problems faced by encryption algorithms. However, they are very sensitive to channel estimation error, especially at low \ac{SNR} \cite{8509094}. In adaptive transmission-based techniques, the parameters are adjusted/adapted based on the location, channel conditions, and \ac{qos} requirements of the legitimate receiver only. For example, precoding \cite{7562191} and channel shortening filter-based \cite{8292335} techniques provide security at the cost of high \ac{PAPR} \cite{1223551}. Similarly, subcarrier selection-based techniques \cite{8093591} provide security at the cost of spectral efficiency degradation. \Ac{AN}-based techniques are also very effective for ensuring secure communication. In these techniques, an interference signal is added by the trusted node to degrade the performance of a legitimate node without affecting the performance of the legitimate receiver. However, the interference signal may cause an increment in \ac{PAPR} and little power degradation due to the sacrifice of the power resources for noise generation \cite{6516879,6881300}. 

Due to the wireless channel decorrelation property, the attacker cannot decode the data even if it knows the algorithm and tries to apply it based on its channel. However, the work in \cite{zhang2018csisnoop} claimed that under certain assumptions the attacker can acquire the \ac{CSI} between the legitimate nodes, which puts channel-based \ac{PLS} techniques at high risk of failure. For instance, the attacker can reverse engineer the beamforming matrix to acquire the \ac{CSI} between legitimate nodes \cite{zhang2018csisnoop}. The beamforming matrix can be estimated under the assumption that the attacker possesses the \ac{CSI} between the legitimate transmitter and itself, and that it is equipped with the same number of antennas as that of a legitimate transmitter. After estimating \ac{CSI} corresponding to legitimate nodes, it can compromise channel-dependent \ac{PLS} techniques like \ac{AN} as well as key generation-based techniques \cite{6739367}. The security can still be ensured in such cases by enforcing poor channel estimation quality at the attacker for the \ac{CSI} between the legitimate transmitter and the attacker, which motivates the need for pilot security. Securing the pilots protects the propagation environment properties from being extracted at the attacker side. Additionally, pilot security plays a critical role in physical layer authentication techniques. 

\subsubsection{PLS for Pilots}
From the pilot security perspective, few techniques are proposed in the literature. In the line of this direction, in \cite{7343356}, the phases of pilots subcarriers are rotated based on the previously estimated \ac{CSI}. However, the channel is assumed to be known at both communicating nodes before the start of the algorithm. Moreover, the phase is manipulated for a single value which makes it easy to attack such a technique. In \cite{7605496, chang2009training}, 
the security of downlink pilots is provided based on CSI at the transmitter via uplink training. Particularly, \ac{AN} is added in the null space of the channel to degrade the channel estimation capability at the attacker, though it may increase the \ac{PAPR} of the signal. On top of that, power allocation between pilot and noise signal needs to be done intelligently in order to enhance security performance. Similarly, in \cite{9070177}, anti-eavesdropping pilots are designed such that the channel can be estimated at legitimate nodes only. However, the proposed algorithm requires full-duplex communication. In \cite{du2018improving}, the legitimate parties employ secret pilots for channel estimation. In the first step, the first node transmits the secret pilot to another node. Afterwards, the receiving node sends the received signal to the transmitting node by using amplify and forward strategy. The second node follows the same procedure as that of the first node. This approach requires extra overhead to share feedback. Although authors in \cite{9095399} exploit the uniqueness of channel responses between different nodes in order to degrade the channel estimation at the attacker node and showed the effect in terms of \ac{BER}, they did not consider the sign ambiguity issue when taking the square root of the channel, which in the practical case leads to huge \ac{BER} due to the erroneous channel estimation.

\subsection{Our Contributions}
In order to address the above-mentioned challenges, we propose novel schemes that can provide security for data, pilots, or joint data and pilots. The design of the proposed schemes is based on the decomposition of the channel into all-pass and minimum-phase channels and exploiting the property of decomposed channel to provide security.

The main contributions of the proposed work are as follows:
\begin{itemize}
\item We propose two novel minimum-phase all-pass channel decomposition-based \ac{PLS} schemes for OFDM in rich scattering channels, where the first method secures the data and the latter secures the pilots. The proposed data security algorithm ensures strong security compared to conventional schemes. Furthermore, it preserves the requirements of the legitimate user without trading off security with overall performance. On the other hand, the proposed pilot security algorithm destroys the ability of the eavesdropper in estimating its channel; thus, protecting CSI, sensing, and radio environment mapping information.

\item To the best of our knowledge, the proposed schemes are the first work that enables flexible adaptive security based on security needs. Specifically, the security of data, pilot, or both can be selected based on the security requirements of the applications. Additionally, the proposed novel security scheme is robust against spatially correlated eavesdroppers located near legitimate nodes.

\item Our novel security methods focusing on both data security and pilot exploit only the all-pass component of the channel which has unit power property. Therefore, we provide security without any power constraint such as \ac{PAPR}. Whereas, the conventional PLS techniques such as artificial noise, zero forcing, and channel shortening provide security at the cost of changing the power distribution of the transmitted signal. These changes create high power peaks in time (i.e., PAPR) and frequency (exceeding spectrum mask limits). 

\item The secrecy of the proposed data security algorithm is evaluated under correlated and uncorrelated eavesdropping channels via closed-form \ac{BER}. Whereas for pilot security, we analyzed the \ac{NMSE} performance of the estimated channel. The simulations very well agree with the analytical formulas emphasizing the effectiveness of the proposed algorithms.

\end{itemize}
\subsection{Organization and Notation}
The rest of this paper is organized as follows. Section \ref{sec:system-model} depicts the system model and discusses OFDM preliminaries. Section \ref{Sec:Proposed algorithm} firstly explains the channel decomposition and then presents the proposed data and pilot security algorithms. The numerical analysis of the proposed schemes is given in Section \ref{Sec:Numerical Analysis} followed by performance and simulation analysis in Section \ref{sec:simulation}. Finally, Section \ref{sec:conclusion} concludes the paper.

Bold uppercase $\mathbf{A}$, bold lowercase $\mathbf{a}$, and unbold letters $A,a$ are used to denote matrices, column vectors, and scalar values, respectively. $(\cdot)^H$, $(\cdot)^T$, and $(\cdot)^{-1}$ denote the Hermitian, transpose, and inverse. $|\cdot|$ denotes the Euclidean norm, E$[\cdot]$ denotes the expectation operator, and var$(\cdot)$ denotes the variance operator. $\mathbb{C}^{{M\times N}}$ denotes the space of $M\times N$ complex-valued matrices. Symbol $j$ represents the imaginary unit of complex numbers with $j^2=-1$.

\section{System model and preliminaries}\label{sec:system-model}
\begin{figure}[t]
	\centering
	\includegraphics[width=0.90\columnwidth]{ 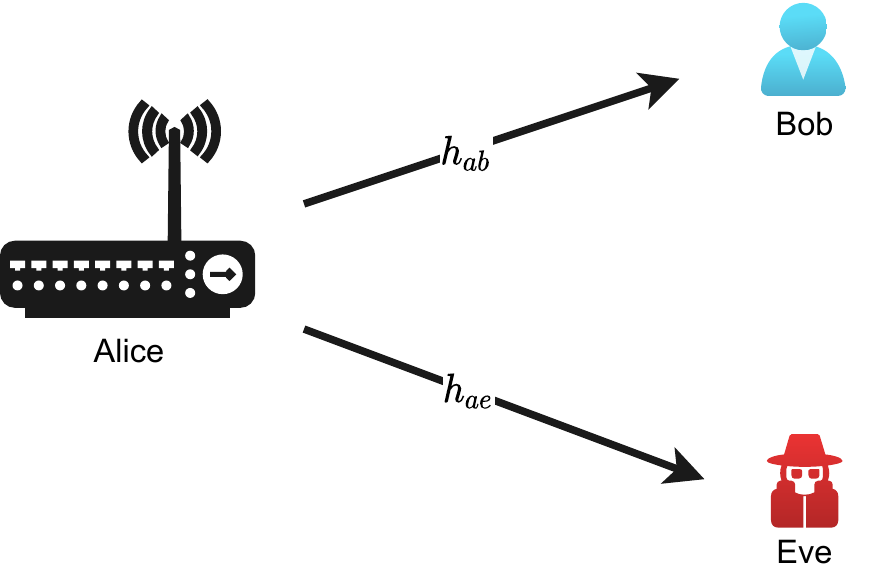}
    	\footnotesize\caption{ System model where Alice and Bob are communicating over rich scattering channel with the existence of Eve.}
	\label{fig:System-model}
\end{figure}

As demonstrated in Fig. \ref{fig:System-model}, a \ac{SISO} \ac{OFDM} system is considered that consists of a legitimate transmitter (Alice, \{a\}), legitimate receiver (Bob, \{b\}),
and passive eavesdropper (Eve, \{e\}) that is trying to intercept the transmission between Alice and Bob, where each node is equipped with a single antenna. The channels observed at Alice $\mathbf{h}_{ba} \in \mathbb{C}^{L \times 1}\sim \mathcal{CN}(0,\sigma_b)$, Bob $\mathbf{h}_{ab} \in \mathbb{C}^{L \times 1}\sim \mathcal{CN}(0,\sigma_a)$, and Eve $\mathbf{h}_{ae} \in \mathbb{C}^{L \times 1}\sim \mathcal{CN}(0,\sigma_e)$ are considered as multi-path slowly varying channels with $L$ exponentially decaying taps having Rayleigh fading distribution. Moreover, due to the channel reciprocity assumption, the channel between Alice-Bob $\mathbf{h}_{ab}$ can be estimated from the channel between Bob-Alice $\mathbf{h}_{ba}$, where $\mathbf{h}_{ab}=\mathbf{h}_{ba}^T$ \cite{Recip}. In addition, as the wireless channel varies due to environment richness and locations of nodes, the channel experienced by Bob and Eve is assumed to be independent. Furthermore, it is also assumed that Alice has no information about the channel of Eve because of its passive operation.

\ac{OFDM} system is adopted for the communication, where $N$ complex data symbols in frequency domain $\mathbf{{X}}= \begin{bmatrix} X(0) & X(1) &...& X(N-1)\end{bmatrix}^{\textrm T} $ are converted to time-domain $\mathbf{x}=\begin{bmatrix} x(0) & x(1) &...& x(N-1)\end{bmatrix}^{\textrm T} $ using \ac{IFFT} to form one \ac{OFDM} symbol as 
\begin{equation}
    {x}(n) = \frac{1}{N}\sum_{k=0}^{N-1}{X}(k)e^{j2\pi nk/N}.
\end{equation}
To combat the \ac{ISI}, a \ac{CP} is appended to $\mathbf{x}$ before transmission. Finally, the signal is transmitted through the wireless channel and reaches the legitimate receiver (Bob) and illegitimate receiver (Eve).

The wireless channel is represented by its \ac{CIR}, which is given as
\begin{equation}
    h_{\Lambda}(t,\tau) = \sum_{l=0}^{L-1}h_l(t)\delta(\tau-\tau_l),
\end{equation}
where $\Lambda \in \{ab,ba,ae\}, $ $h_l(t)$ and $\tau_l$ are the complex channel gain and the delay of the $l$-th path at time $t$, respectively. $h_l(t)$ is assumed to have Gaussian distribution with zero-mean. $L$ is the total number of effective channel taps and $\delta(\cdot)$ is the Kronecker delta function. The \ac{CFR} is then expressed as
\begin{equation}
    H_{\Lambda}(t,f) = \int^{+\infty}_{-\infty}h_{\Lambda}(t,\tau)e^{-j2\pi f\tau}d\tau.
\end{equation}
Assuming that the channel is time-invariant during one \ac{OFDM} symbol period $T_s$, and that the frequency spacing is $\Delta f_c$, the \ac{CIR} and the \ac{CFR} can be respectively represented as 
\begin{equation*}
    h_{\Lambda}(n) = h_{\Lambda}(nT_s,\tau),~ H_{\Lambda}(k) = H_{\Lambda}(nT_s,k\Delta f_c) .
\end{equation*}

At the receiver, the \ac{CP} is discarded first, and then the \ac{FFT} process is applied. The received $k$-th symbol is found as
\begin{equation}
    Y_{\Lambda}(k) = H_{\Lambda}(k)X_{\Lambda}(k)+W(k),
\end{equation}
where $W(k)$ is the zero-mean \ac{AWGN} with variance $\sigma^2$ at the $k$-th subcarrier. It is assumed that the sampling rate satisfies the Nyquist criteria.

Given that the \ac{CFR} is estimated using the known pilots after receiving the signal, let $k_p$ be the $p$-th index where the pilot is inserted in the data signal $\mathbf{X}$, then the estimated \ac{CFR} is considered as
\begin{equation}
    \tilde{H}_{\Lambda}(k_p) = \frac{Y(k_p)}{X(k_p)} = H_{\Lambda}(k_p)+\tilde{Z}(k_p),
     \label{Hp}
\end{equation}
where $\tilde{W}(k_p)$ denotes the noise term. To get an estimation over all $N$ subcarriers, $\tilde{H}$ is interpolated and the final estimated \ac{CFR} is found.


\begin{figure*}[!t]
    \begin{center}
    \subfloat[\footnotesize Overall Channel.]{\label{convPerf:1}\includegraphics[width=58mm]{ 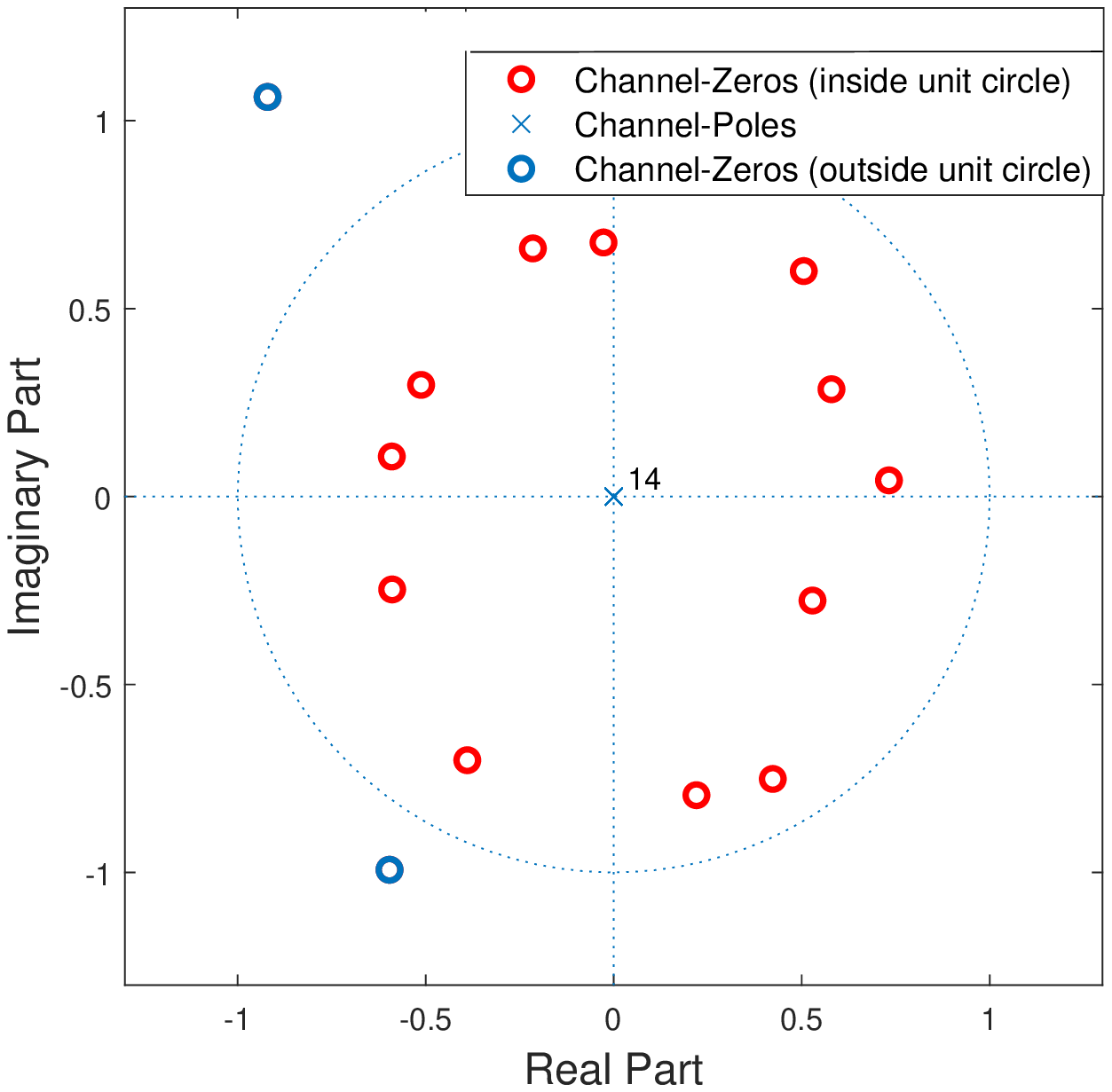}}
    \subfloat[\footnotesize All-pass channel.]{\label{convPerf:2}\includegraphics[width=58mm]{ 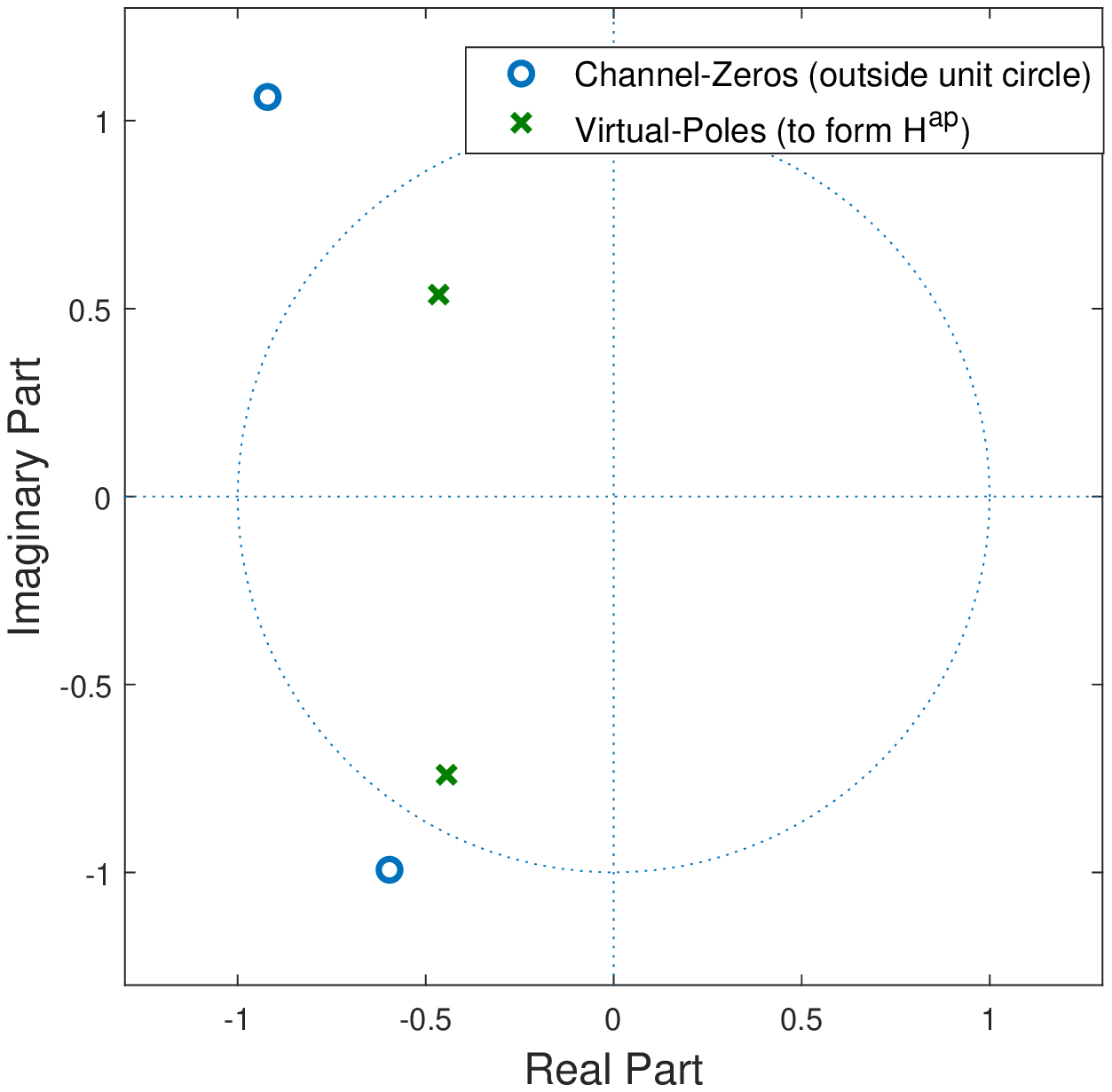}}
    \subfloat[\footnotesize Minimum-phase channel.]{\label{convPerf:3}\includegraphics[width=58mm]{ 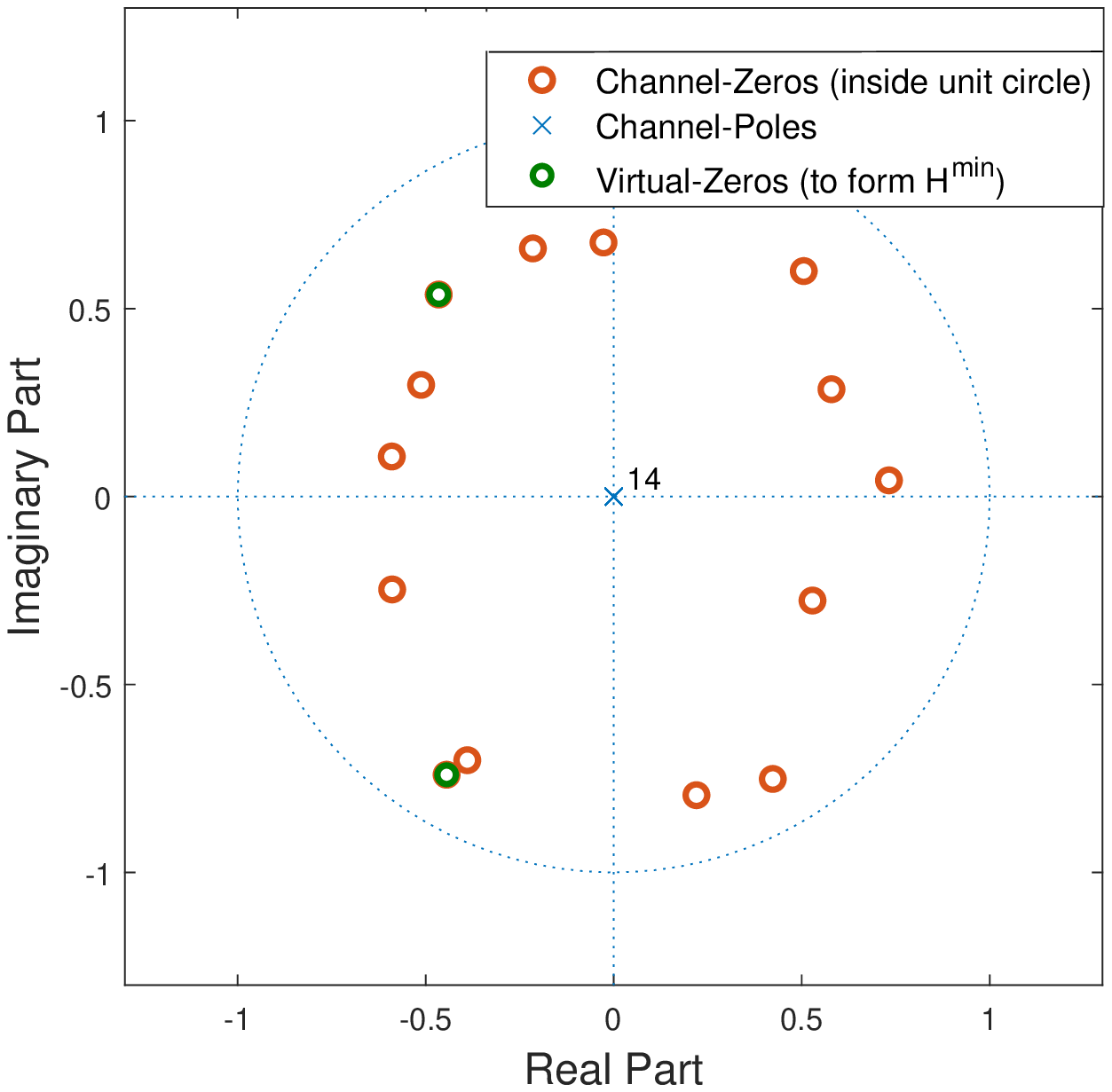}}
    \\
    \end{center}
    \centering
    \footnotesize\caption{ The zero-pole diagram of the minimum-phase all-pass decomposition of a wireless channel.}
    \label{fig:Channel-decomposition}
\end{figure*}


\section{Proposed Algorithm}\label{Sec:Proposed algorithm}
The increase in the number of wireless communication-based applications with varying requirements motivates the need for adaptive and flexible security designs \cite{9336039}. Inspired by this motivation, in this section, we firstly present the channel decomposition concept, and then we propose novel algorithms that are capable of providing adaptive and flexible security. Particularly, in the case of a very high level of security, the security of the pilot and data is provided using the proposed algorithm. Otherwise, the security of data or pilot is provided based on the security requirements.

\subsection{Minimum-phase All-pass Channel Decomposition}\label{Subsec:Channel decomp}
\begin{figure*}[t]
    \centering
    \includegraphics[scale=0.65]{ 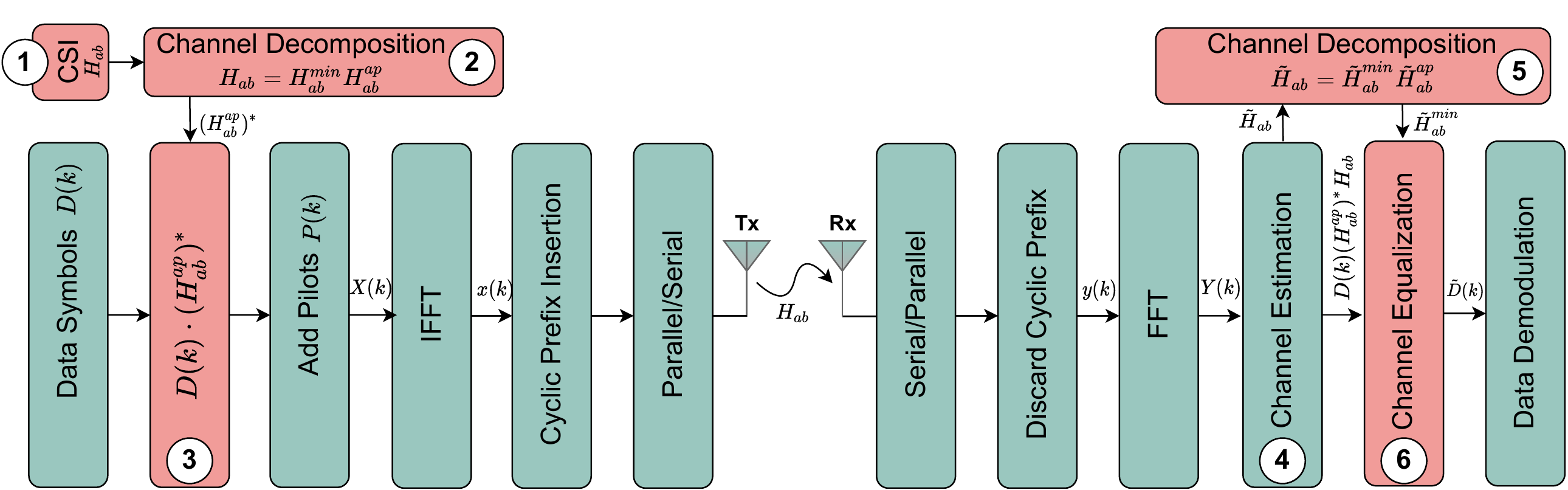}
    \footnotesize\caption{ The block diagram showing the main steps of the proposed data security algorithm.}
    \label{fig:Data security}
\end{figure*}

Wireless channel systems are causal because they are real-time systems, where the samples belong only to the present or past. Additionally, the \ac{CIR} of a wireless channel can be represented by a \ac{FIR} filter, and thus it is stable \cite{oppenheim}. Consequently, a stable and causal system with system function $H_{\Lambda}(z)$ would have all poles inside its unit circle; however, the zeros are free to wander outside. Let $H_{\Lambda}^1(z)$ be the system function with all zeros and poles inside the unit circle, and let the zeros outside be at $1/p_k$. This implies that we can decompose such a system into two components as
\begin{equation}
H_{\Lambda}(z)=\underbrace{\left(H_{\Lambda}^1(z) \prod_{k=1}^{q}\left(1-p_{k}^{*} z^{-1}\right)\right)}_{H_{\Lambda}^{\min }(z)} \overbrace{\prod_{k=1}^{q}\left(\frac{z^{-1}-p_{k}}{1-p_{k}^{*} z^{-1}}\right)}^{H_{\Lambda}^{\mathrm{ap}}(z)},
\label{equ:factorization}
\end{equation}
where $H_{\Lambda}^{\mathrm{min}}(z)$ and $H_{\Lambda}^{\mathrm{ap}}(z)$ are defined as the minimum-phase and all-pass components of $H_{\Lambda}(z)$, respectively. For instance, Fig. \ref{fig:Channel-decomposition} illustrates the zero-pole diagram of the minimum-phase all-pass decomposition of a random channel. As seen in Fig. \ref{fig:Channel-decomposition}(a), the overall channel contains zeros inside and outside the unit circle where the poles are centered at the origin. As shown in Fig. \ref{fig:Channel-decomposition}(b), for the all-pass channel only the zeros outside the unit circle are considered along with virtual poles added at the inverse of the zeros' location to cancel out the attenuation effect, thus passing all frequencies as the name stands. To compensate the effect of these virtual poles, zeros are added on top of them having a system with all zeros inside the unit circle (i.e., minimum-phase system) as illustrated in Fig. \ref{fig:Channel-decomposition}(c).

The resulting components have various properties; for instance, in terms of the magnitude response $|H(e^{j\omega})|$, the factorization in \eqref{equ:factorization} implies that $|H_{min}(e^{j\omega})| = |H(e^{j\omega})|$ and $|H_{ap}(e^{j\omega})| = 1$. These properties of the decomposed channel will be exploited to provide security for both data and pilots.

\subsection{Proposed Data Security Method}\label{Subsec:Data security}
This subsection presents the details of the proposed algorithm for providing data security. The designed algorithm is based on a novel precoder that exploits the components of the channel separately, instead of using the full channel as in conventional security algorithms \cite{7467419}. As explained in Subsection \ref{Subsec:Channel decomp}, the proposed method uses only the conjugate of all-pass $H_{ap}^*(e^{j\omega})$ component of the channel for precoding. Therefore, it will not enhance the \ac{PAPR} \cite{4543070}. Additionally, it provides an effective solution to ensure secure communication against eavesdropping.

\begin{figure*}[t]
    \centering
    \includegraphics[scale=0.65]{ 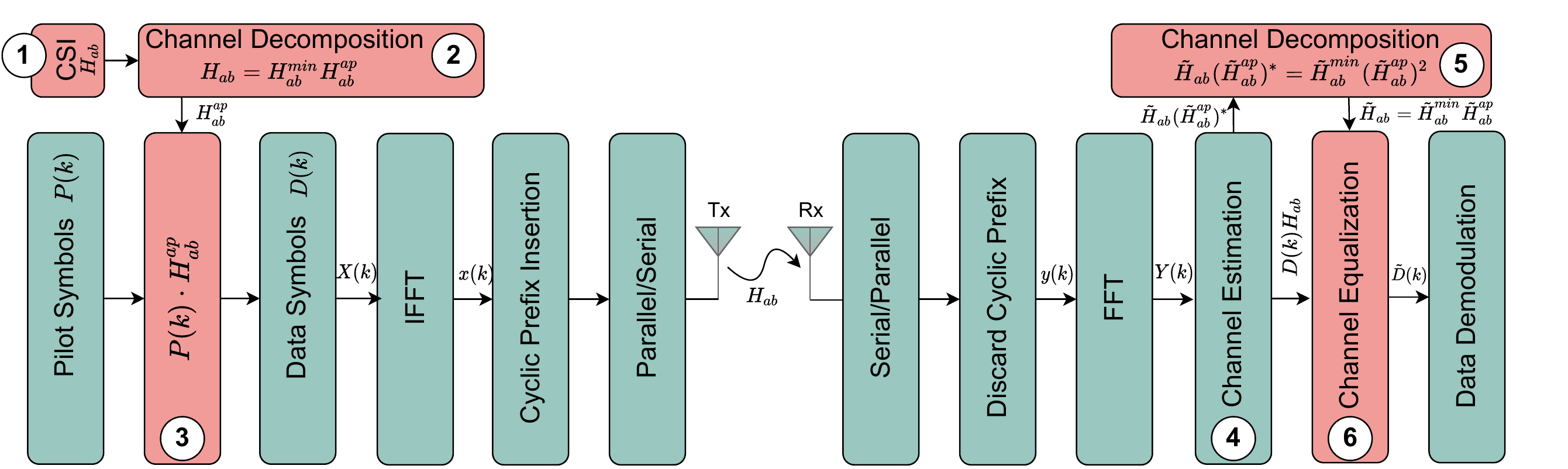}
       \footnotesize\caption{ The block diagram showing the main steps of the proposed pilot security algorithm.}

    \label{fig:Pilot security}
\end{figure*}

Fig. \ref{fig:Data security} illustrates the block diagram of the proposed data security algorithm, where its basic steps are described as follows:
\begin{enumerate}
 \item Bob sends pilot signal ${P}$ to Alice to estimate $H_{ba}$, where due to channel reciprocity $H_{ba}=H_{ab}$. Thus, we assume that \ac{CSI} is available at Alice. 
 \item Alice decomposes the \ac{CFR} $H_{ba}$ into all-pass $H_{ab}^{\mathrm{ap}}$ and minimum-phase $H_{ab}^{\mathrm{min}}$ components, as explained in Subsection \ref{Subsec:Channel decomp}.
 \item Alice multiplies the data subcarriers $X(k_d)$ at indices $k_d$ with the conjugate of all-pass components of the channel (${H_{ab}^{\mathrm{ap}}}^*$), while the pilots $X(k_p)$ at $k_p$ indices are intact. Then, the transmitted signal by Alice can be expressed as
	\begin{equation}
    X(k) =
    \begin{cases}
      {H_{ab}^{\mathrm{ap}}}^*(k)D(k) &; k \in k_d\\
      P(k) &; k \in k_p.\\
   \end{cases} 
   \end{equation}
\item The received signal at Bob can be given as:
		\begin{equation}
		\begin{aligned}
	Y_{ab}(k)=
    \begin{cases}
      H_{ab}(k){H_{ab}^{\mathrm{ap}}}^*(k)D(k)+W_{ab}(k) &; k \in k_d\\
      H_{ab}(k)P(k)+W_{ab}(k) &; k \in k_p.\\
   \end{cases} 
   \end{aligned}
   \end{equation}
 Using the pilots $P(k)$ at $k_p$, the \ac{CFR} $\tilde{H}_{ab}(k)$ is estimated as described in \eqref{Hp}. 
 
\item Applying channel decomposition to the estimated channel as in Subsection \ref{Subsec:Channel decomp}, we obtain $\tilde{H}_{ab}(k)=\tilde{H}_{ab}^{\mathrm{min}}(k)\tilde{H}_{ab}^{\mathrm{ap}}(k)$. The data subcarriers of the received signal by Bob is given as
\begin{equation}
\begin{aligned}
     Y_{ab}(k) &= H_{ab}^{\mathrm{min}}(k)H_{ab}^{\mathrm{ap}}(k){H_{ab}^{\mathrm{ap}}}^*(k)D(k)+W_{ab}(k)\\
     &=H_{ab}^{\mathrm{min}}(k)D(k)+W_{ab}(k);~ k\in k_d.
\end{aligned}
\label{equ:rec_ab_2}
\end{equation}
\item Using the results of step 5, Bob equalizes $H_{ab}^{\mathrm{min}}$ to decode the data as 
\begin{equation}
\begin{aligned}
     \hat{X}_{Bob}(k) &= \frac{Y_{ab}(k)}{\tilde{H}_{ab}^{\mathrm{min}}(k)}\\
     &=\frac{H_{ab}^{\mathrm{min}}(k)D(k)+W_{ab}(k)}{\tilde{H}_{ab}^{\mathrm{min}}(k)}\\
     &= D(k)+\tilde{W}_{ab}(k);~ k\in k_d,
\end{aligned}
\label{dateqe}
\end{equation}
where $\tilde{W}_{ab}(k)= W_{ab}(k)/\tilde{H}_{ab}^{\mathrm{min}}(k)$ and $H_{ab}^{\mathrm{min}}(k)=\tilde{H}_{ab}^{\mathrm{min}}(k)$ in case of perfect channel estimation. 
\end{enumerate}

The received signal at Eve can be given by
	\begin{equation}
	\begin{aligned}
	Y_{ae}(k)=
    \begin{cases}
      H_{ae}(k){H_{ab}^{\mathrm{ap}}}^*(k)D(k)+W_{ae}(k) &; k \in k_d\\
      H_{ae}(k)P(k)+W_{ae}(k) &; k \in k_p.\\
   \end{cases} 
   \end{aligned}
   \end{equation}
Applying the similar technique at Eve, the final signal at Eve can be given as 
\begin{equation}
\begin{aligned}
     \hat{X}_{eve}(k) &= \frac{Y_{ae}(k)}{\tilde{H}_{ae}(k)}\\
     &=\frac{H_{ae}(k){H_{ab}^{\mathrm{ap}}}^*(k)D(k)+W_{ae}(k)} {\tilde{H}_{ae}(k)}\\
     &= {H_{ab}^{\mathrm{ap}}}^*(k)D(k)+\tilde{W}_{ae}(k);~ k\in k_d,
\end{aligned}
\label{equ:eve-data}
\end{equation}
where $\tilde{W}_{ae}(k)= W_{ae}(k)/\tilde{H}_{ae}(k)$ and $H_{ae}(k)=\tilde{H}_{ae}$ in case of perfect channel estimation. 

As seen from \eqref{equ:eve-data}, Eve will not be able to decode the data even if it perfectly estimates its channel. This is due to the unknown randomness caused by the term ${H_{ab}^{\mathrm{ap}}}^*$ which is uncorrelated with its channel.\footnote{Note that due to channel decorrelation in rich scattering environment between $\mathbf{H}_{ab}$ and $\mathbf{H}_{ae}$, Eve will not be able to estimate and remove the effect of ${H_{ab}^{\mathrm{ap}}}^*(k)$.}

\subsection{Proposed Pilot Security Method}\label{Subsec:Pilot Security}
This subsection presents the details of the proposed algorithm for providing pilot security. Similar to data security, the proposed pilot security algorithm exploits the components of the channel separately. It ensures that only a legitimate receiver will be able to estimate the channel while Eve will not able to learn the channel or environment without affecting \ac{PAPR} as in \cite{7605496}. Additionally, the proposed algorithm is also suitable for the security of feedbacks, hardware impairments, and hardware-based authentication. Furthermore, the eavesdropper will not be able to extract information of precoder corresponding to the channel of legitimate nodes from the received signal and thus will not be able to launch attacks to learn \ac{CSI} corresponding to legitimate node \cite{zhang2018csisnoop}.

Fig. \ref{fig:Pilot security} illustrates the block diagram of the proposed pilot security algorithm, where its basic steps are described as follows:
	\begin{enumerate}
	\item Bob sends pilot signal ${P}$ to Alice to estimate $H_{ba}$, where due to channel reciprocity $H_{ba}=H_{ab}$. Thus we assume that \ac{CSI} is available at Alice. 
	\item Alice decomposes the \ac{CFR} $H_{ba}$ into all-pass $H_{ab}^{\mathrm{ap}}$ and minimum-phase $H_{ab}^{\mathrm{min}}$ components as explained in Subsection \ref{Subsec:Channel decomp}.
	\item Alice multiplies the pilots subcarriers $X(k_p)$ at indices $k_p$ with the all-pass components of the channel ${H_{ab}^{\mathrm{ap}}}$, while the data subcarriers $X(k_d)$ at $k_d$ indices are intact. Then, the transmitted signal by Alice can be expressed as
	
	\begin{equation}
    X(k) =
    \begin{cases}
      D(k) &; k \in k_d\\
      {H_{ab}^{\mathrm{ap}}}(k)P(k) &; k \in k_p.\\
   \end{cases} 
   \end{equation}

	\item The received signal at Bob can be given as
		\begin{equation}
		\begin{aligned}
	Y_{ab}(k)=
    \begin{cases}
     H_{ab}(k)D(k)+W_{ab}(k) &; k \in k_d\\
     H_{ab}(k){H_{ab}^{\mathrm{ap}}}(k)
     P(k)+W_{ab}(k) &; k \in k_p.\\
   \end{cases} 
   \end{aligned}
   \end{equation}
 Using the pilots $P(k)$ at $k_p$, the precoded \ac{CFR} $\tilde{H}_{abp}(k)=H_{ab}(k){H_{ab}^{\mathrm{ap}}}(k)$ is estimated as described in \eqref{Hp}. 
 
 \item In order to find $\tilde{H}_{ab}(k)$ from the estimated $\tilde{H}_{abp}(k)$, channel decomposition is applied to the estimated precoded channel as explained in Subsection \ref{Subsec:Channel decomp} as follows: $\tilde{H}_{abp}(k)=\tilde{H}_{ab}^{\mathrm{min}}(k) 
 (\tilde{H}_{ab}^{\mathrm{ap}}(k))^2$, where $\tilde{H}_{ab}^{\mathrm{min}}(k)$ is minimum-phase component while $(\tilde{H}_{ab}^{\mathrm{ap}}(k))^2$ is all-pass component of $\tilde{H}_{abp}(k)$.

\item The estimated channel at Bob can be calculated as: 
 $\tilde{H}_{ab}(k)=\tilde{H}_{ab}^{\mathrm{min}}(k)\sqrt{(\tilde{H}_{ab}^{\mathrm{ap}}(k))^2}$.\footnote{Note that $\sqrt{(\tilde{H}_{ab}^{\mathrm{ap}})^2}=\pm \tilde{H}_{ab}^{\mathrm{ap}}$. Therefore, in order to estimate the sign of estimated channel, we exploit the correlation between the channel subcarriers which ensure a smooth transition between the value of one subcarrier to another. Thus, we solved the sign ambiguity compared to the work in \cite{9095399} which didn't consider such issue.} 
\end{enumerate}

At Eve side, the received signal is given by
	\begin{equation}
	\begin{aligned}
	Y_{ae}(k)=
    \begin{cases}
     H_{ae}(k)D(k)+W_{ae}(k) &; k \in k_d\\
     H_{ae}(k){H_{ab}^{\mathrm{ap}}}(k)
     P(k)+W_{ae}(k) &; k \in k_p.\\
   \end{cases} 
   \end{aligned}
   \label{equ:eve-pilot}
   \end{equation}
Using the pilots $P(k)$ at $k_p$, the precoded \ac{CFR} $\tilde{H}_{abe}(k)=H_{ae}(k){H_{ab}^{\mathrm{ap}}}(k)$ is estimated as described in \eqref{Hp}. As seen from \eqref{equ:eve-pilot}, eavesdropper will not be able to correctly estimates the channel because of unknown randomness caused by all-pass component (${H_{ab}^{\mathrm{ap}}}(k)$) of legitimate channel in \eqref{equ:eve-pilot}. Hence, it cannot estimate the channel and learn the environment.

\subsection{Joint Pilot \& Data Security}\label{Subsec:Joint Security}
This subsection presents the details of the proposed algorithm for providing joint data and pilot security. Particularly, in case of a very high-security risk, there is a need of securing both pilot and data to provide a very high level of security. Thus, the attacker will not able to learn the channel, environment, and data. Here, the idea is to exploit the proposed algorithm presented in Subsections \ref{Subsec:Data security} and \ref{Subsec:Pilot Security} to ensure both pilot and data security. 
The transmitted signal by Alice after applying both data and pilot security can be expressed as

	\begin{equation}
    X(k) =
    \begin{cases}
      {H_{ab}^{\mathrm{ap}}}^*(k)D(k) &; k \in k_d\\
      {H_{ab}^{\mathrm{ap}}}(k)P(k) &; k \in k_p.\\
   \end{cases} 
   \end{equation}

The received signal at Bob can be given as
		\begin{equation}
		\begin{aligned}
		Y_{ab}(k)=
    \begin{cases}
      H_{ab}(k){H_{ab}^{\mathrm{ap}}}^*(k)D(k)+W_{ab}(k) &; k \in k_d\\
      H_{ab}(k){H_{ab}^{\mathrm{ap}}}(k)
       P(k)+W_{ab}(k) &; k \in k_p.\\
   \end{cases} 
   \end{aligned}
   \end{equation}
Using the pilots $P(k)$ at $k_p$, the precoded \ac{CFR} $\tilde{H}_{abp}(k)=H_{ab}(k){H_{ab}^{\mathrm{ap}}}(k)$ is estimated as described in \eqref{Hp}. Afterwards, $\tilde{H}_{ab}(k)$ is estimated as $\tilde{H}_{ab}(k)=\tilde{H}_{ab}^{\mathrm{min}}(k)\sqrt{(\tilde{H}_{ab}^{\mathrm{ap}})^2}$.
Finally, Bob will decode the data similar to \eqref{dateqe}.

On the other hand, the received signal at Eve can be given as:
		\begin{equation}
		\begin{aligned}
       Y_{ae}(k)=
    \begin{cases}
      H_{ae}(k){H_{ab}^{\mathrm{ap}}}^*(k)D(k)+W_{ae}(k) &; k \in k_d\\
      H_{ae}(k){H_{ab}^{\mathrm{ap}}}(k)
       P(k)+W_{ae}(k) &; k \in k_p.\\
   \end{cases} 
   \end{aligned}
  \label{jointeve}
   \end{equation}
  It should be noted from \eqref{jointeve} that Eve will neither be able to estimate its channel nor the data due to the randomness caused by ${H_{ab}^{\mathrm{ap}}}^*(k)$ and ${H_{ab}^{\mathrm{ap}}}(k)$ in $D(k)$ and in $P(k)$, respectively. Thus, providing a two-level security that is suitable for critical applications. 
  
\section{Numerical Analysis}\label{Sec:Numerical Analysis}
In this section, we analyze the BER performance for Bob and Eve under correlated and uncorrelated eavesdropping channels to investigate the data security algorithm. Afterwards, we derive the \ac{MMSE} of the estimated channel when the pilot security algorithm is applied.

\subsection{Data Security: BER-based Secrecy Gap}\label{Subsec:Data security analysis}
To emphasize the performance of the data security method, we compare the \ac{BER} performance gap between Bob and Eve. In this subsection, we analyze the \ac{BEP} under correlated and uncorrelated eavesdropping channels.
\subsubsection{Uncorrelated Bob-Eve Channel}

One elaborate method to suppress the effect of $W(k)$ when estimating the channel is the \ac{MMSE} estimation \cite{MMSE_OFDM}. After estimating the channel $\tilde{H}$, the data at index $k_d$ is given in \eqref{equ:rec_ab_2} and expressed by
\begin{equation}
     Y_{ab}(k)=H_{ab}^{\mathrm{min}}(k)D(k)+W(k);~ k\in k_d.
\end{equation}

Therefore, for a normalized power data symbols (i.e., $\operatorname{E}[|D(k)|^2]=1$) the \ac{SNR} of the received signal is given by
\begin{equation}
\begin{aligned}
     \gamma_{ab}&\triangleq\frac{\operatorname{E}[|H_{ab}^{\mathrm{min}}(k)D(k)|^2]}{M\operatorname{E}[|W(k)|^2]}=\frac{\operatorname{E}[|H_{ab}^{\mathrm{min}}(k)|^2]\operatorname{E}[|D(k)|^2]}{M\operatorname{E}[|W(k)|^2]}\\
     &=\frac{\operatorname{E}[|H_{ab}^{\mathrm{min}}(k)|^2]}{M\sigma^2},
\end{aligned}
\end{equation}
where $M$ denotes the number of bits represented by each symbol.

As demonstrated in Subsection \ref{Subsec:Channel decomp}, the minimum-phase component shares the same power with the overall channel response itself $|H_{ab}^{\mathrm{min}}(k)|^2 = |H_{ab}(k)|^2$. Thus, the \ac{SNR} can be written as
\begin{equation}
     \gamma_{ab}=\frac{\operatorname{E}[|H_{ab}(k)|^2]}{M\sigma^2}.
\end{equation}

The average \ac{BEP} $P_{\Lambda}(e)$ then is given as function of \ac{SNR} $\gamma_\Lambda$ and the correlation coefficients $\rho_\Lambda^1$ and $\rho_\Lambda^2$ between the actual and the estimated channel responses \cite{OFDM_performance} by:
\begin{equation}
\begin{aligned}
P_{\Lambda}(e)=\frac{1}{2}&\left[1-\frac{1}{2} \frac{\frac{\left(\rho_\Lambda^1+\rho_\Lambda^2\right)}{\sqrt{2}}}{\sqrt{1+\frac{1}{2 {\bar{\gamma}}_{\Lambda}}-\frac{\left(\rho_\Lambda^1-\rho_\Lambda^2\right)^{2}}{2}}}\right. \\
&\left.-\frac{1}{2} \frac{\frac{\left(\rho_\Lambda^1-\rho_\Lambda^2\right)}{\sqrt{2}}}{\sqrt{1+\frac{1}{2 \bar{\gamma}_{\Lambda}}-\frac{\left(\rho_\Lambda^1+\rho_\Lambda^2\right)^{2}}{2}}}\right],
\end{aligned}
\label{equ:BER_imperfect}
\end{equation}
where $\Lambda=\{ab,ae\}$.

In case of perfect channel estimation, we have $\operatorname{E}[|H_{ab}-\tilde{H}_{ab}|^2]=0$, $\rho_{ab}^1 = 1$ and $\rho_{ab}^2 =0$, and the \ac{BEP} performance will become the lower bound performance of \eqref{equ:BER_imperfect}, and it is given as   
\begin{equation}
P_{ab}(E)=\frac{1}{2}\left[1- \frac{1}{\sqrt{1+\frac{1}{ \bar{\gamma}_{ab}}}}\right].
\label{equ:BER_perfect}
\end{equation}

\begin{figure} [t]
\centering
\begin{subfigure}[b]{0.45\textwidth}
   \includegraphics[width=1\linewidth]{ 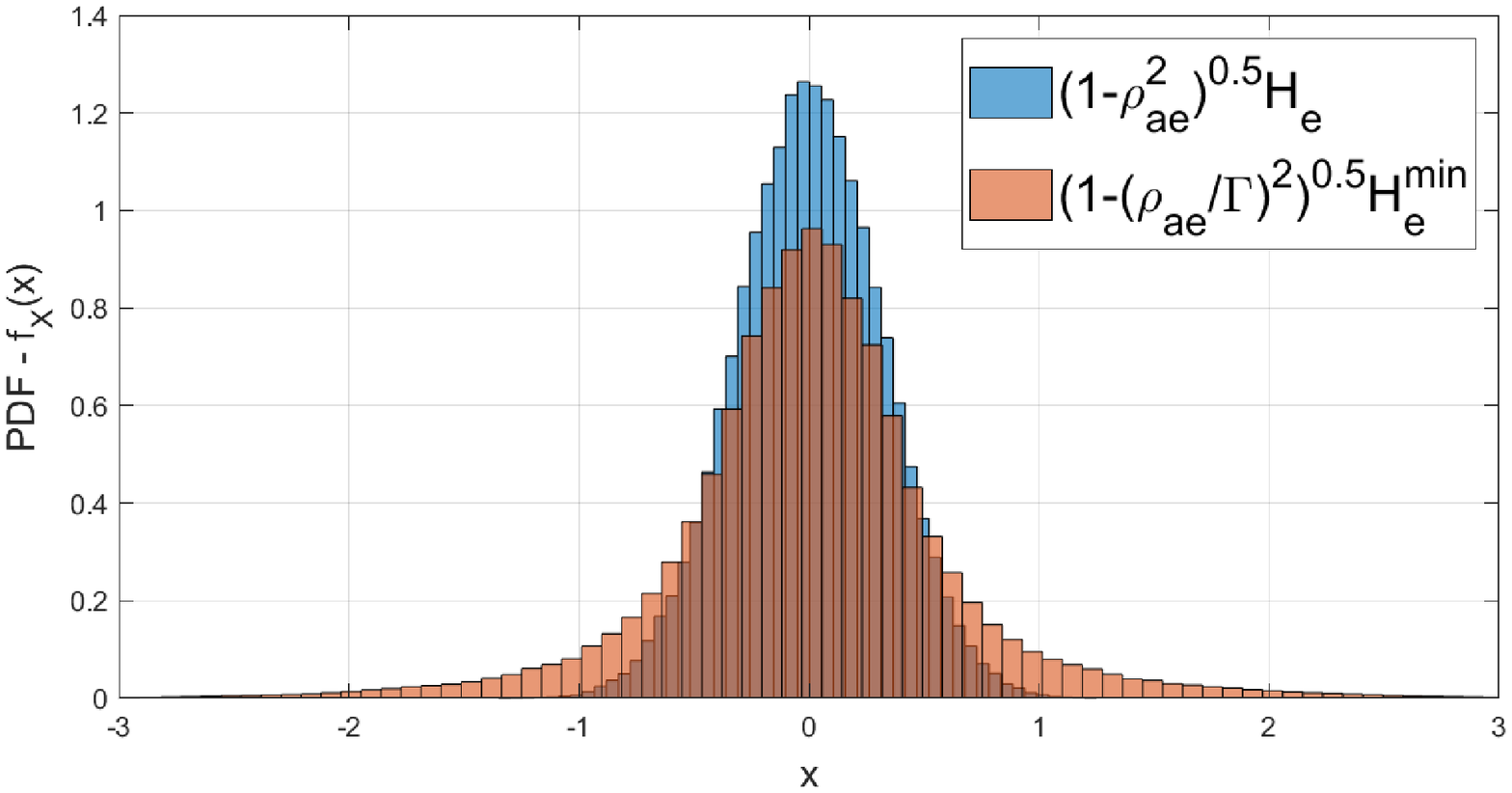}
   \caption{ \footnotesize Real part.}
\end{subfigure}
\begin{subfigure}[b]{0.45\textwidth}
   \includegraphics[width=1\linewidth]{ 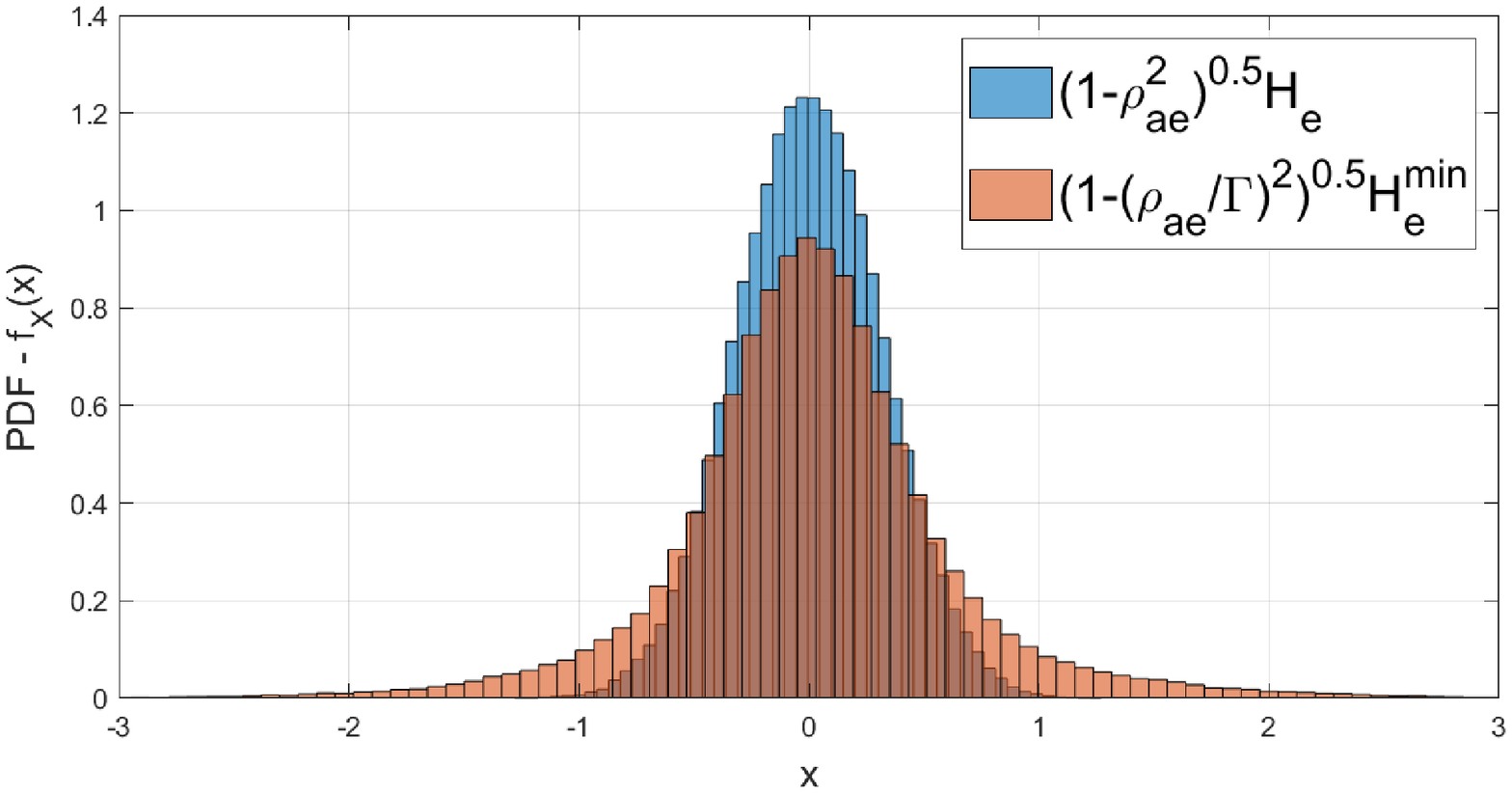}
   \caption{\footnotesize Imaginary part.}
\end{subfigure}
\footnotesize\caption{ The distribution of uncorrelated part of Eve's channel for $\rho_{ae}=0.9$.}\label{fig:PDF}
\end{figure}

\subsubsection{Correlated Bob-Eve Channel}
To further evaluate the performance of the proposed scheme and emphasize its reliability, we consider the effects of the eavesdropper location with respect to the legitimate user. For that, we consider a correlated eavesdropping channel scenario, where it is assumed that Eve is located near Bob. We model the correlation between channel coefficients \cite{ferdinand2013physical}/ as
\begin{equation}
    H_{ae} = \rho_{ae}H_{ab}+\sqrt{1-\rho_{ae}^2}H_{e},
    \label{equ:corr_chan}
\end{equation}
where $H_{e}$ is i.i.d. $\sim \mathcal{CN}(0,\sigma_a)$ and $\rho_{ae}$ is the correlation function of the legitimate channel gain with the eavesdropping channel given as
\begin{equation}
    \rho_{ae} =\frac{\operatorname{Cov}[{H}_{ab},H_{ae}^*]}{\sqrt{\operatorname{var}({H}_{ab})\operatorname{var}(H_{ae}^*)}}=\frac{\operatorname{E}[{H}_{ab}H_{ae}^*]}{\sigma_a\sigma_e}.
 \label{equ:rho_ae}
\end{equation}
Please refer to Appendix \ref{App:rho}.\hfill$\blacksquare$


In the proposed algorithm, the overall channel is not used, and instead, only one component (i.e., all-pass channel) is exploited. Therefore, even if Eve estimates the channel with high correlation, still it will suffer from the error raised due to the decomposition of its channel. This leads to lower eavesdropping channel correlation and enhanced security performance. Fig. \ref{fig:PDF} shows the distribution of the real and imaginary parts of the uncorrelated part of the channel $H_e$. It is clearly seen that the variance of the uncorrelated term, given blue color, increases after performing the channel decomposition as shown by the orange distribution in Fig. \ref{fig:PDF}. 

\begin{figure}[t]
	\centering
	\includegraphics[scale=0.38]{ 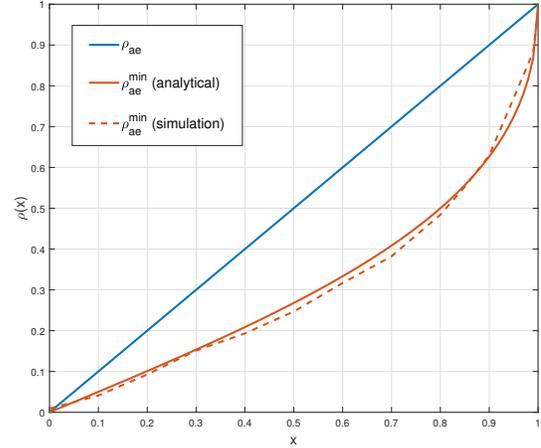}
    	\footnotesize\caption{The channel correlation relationship between the conventional and proposed schemes.}
	\label{fig:Correl}
\end{figure}

Assuming the same channel model as in \eqref{equ:corr_chan} we have
\begin{equation}
 H_{ae} = \rho_{ae}H_{ab}^{\mathrm{min}}H_{ab}^{\mathrm{ap}}+\sqrt{1-\rho_{ae}^2}H_{e}.
\end{equation}

For the channel equalization, only the minimum-phase component of the channel is needed. So, after performing the channel decomposition and normalization, Eve finds
\begin{equation}
\begin{aligned}
    H^{\mathrm{min}}_{ae} &=\rho_{ae}^{\mathrm{min}}H^{\mathrm{min}}_{ab}+\sqrt{1-\left(\rho_{ae}^{\mathrm{min}}\right)^2}H^{\mathrm{min}}_{e}\\
    &=\frac{\rho_{ae}}{\Gamma}H^{\mathrm{min}}_{ab}+\sqrt{1-\left(\frac{\rho_{ae}}{\Gamma}\right)^2}H^{\mathrm{min}}_{e},
\end{aligned}
    \label{equ:corr_min}
\end{equation}
where $\Gamma$ is the correlation attenuation factor. Note that $\Gamma$ satisfies the following constraints
 \begin{equation}
\begin{aligned}
      & \Gamma \sim \sqrt{1-\rho_{ae}^2} ~~~~~~~~~(C1)\\
      &\Gamma \geq 1~\forall \rho_{ae}~~~~~~~~~~~~(C2)\\
      & \Gamma = 1,~ \mathrm{for} ~\rho_{ae}=1~~~(C3) \\
      &0\leq\frac{\rho_{ae}}{\Gamma}\leq1,~\forall \rho_{ae}~~~(C4).
\end{aligned}
\label{equ:const}
\end{equation}
Taking all the constraints given in \eqref{equ:const}, we find that 
\begin{equation}
   \Gamma = 1+\sqrt{1-\rho_{ae}^2}.
    \label{equ:gamma}
\end{equation}
Please refer to Appendix \ref{App:gamma}.\hfill$\blacksquare$

Fig. \ref{fig:Correl} shows the relationship between the correlation of the overall channel of Eve and Alice and the correlation of the minimum-phase component of their channels. As observed from Fig. \ref{fig:Correl}, the correlation factor decreases after performing the channel decomposition causing degradation in the data decoding capability at Eve.

Therefore, the correlation function becomes $\rho_{ae}^{\mathrm{min}} = \rho_{ae}/(1+\sqrt{1-\rho_{ae}^2})<\rho_{ae}$. This result shows another advantage of using the proposed scheme in case of correlated eavesdropping channels. Please refer to Appendix \ref{App:corr_min}.\hfill$\blacksquare$

To evaluate the results above, the \ac{BER} performance is analyzed. we adopt the same \ac{BER} expression given by \eqref{equ:BER_perfect} by including the spatial correlation factor $\rho$ \cite{9095399}. Then, the \ac{BER} at both Bob and Eve can be found using \eqref{equ:BER_imperfect} as
\begin{equation}
P_{\Lambda}(E)=\frac{1}{2}\left[1- \frac{\rho_{\Lambda}}{\sqrt{1+\frac{1}{ \bar{\gamma}_{ab}}}}\right],
\end{equation}
where $\Lambda=\{ab,ae\}$.

\subsection{Pilot Security: Channel NMSE}\label{Subsec:Pilot Security analysis}
Using the pilot scheme as adopted in Subsection \ref{Subsec:Pilot Security}, the \ac{MMSE} estimate of the channel can be obtained as \cite{MMSE_OFDM}

\begin{equation}
    \tilde{H}_{ab}^{\mathrm{MMSE}} = F\tilde{\mathbf{h}}_{ab} = FR_{h_{ab}Y_{ab}}R_{Y_{ab}Y_{ab}}Y^{-1},
     \label{equ:H_MMSE1}
\end{equation}
where 
\begin{equation}
\begin{aligned}
   &R_{h_{ab}Y_{ab}} = \operatorname{E}[h_{ab}Y_{ab}^H]= R_{h_{ab}h_{ab}}F^HP^H\\
   &R_{Y_{ab}Y_{ab}} = \operatorname{E}[Y_{ab}Y_{ab}^H]= PFR_{h_{ab}h_{ab}}F^HP^H+\sigma^2I_N,
\end{aligned}
\label{equ:H_MMSE2}
\end{equation}
where $R_{h_{ab}h_{ab}}$ is the channel autocorrelation matrix, $F$ is the \ac{DFT} matrix, and $I_N$ is the identity matrix. Substituting \eqref{equ:H_MMSE2} in \eqref{equ:H_MMSE1} we find
\begin{equation}
    \tilde{H}_{ab}^{\mathrm{MMSE}} = FR_{h_{ab}h_{ab}}F^HP^H(PFR_{h_{ab}h_{ab}}F^HP^H+\sigma^2I_N)^{-1}Y_{ab}.
     \label{equ:H_MMSE}
\end{equation}

Decomposing the estimated channel $\tilde{H}_{ab}^{\mathrm{MMSE}}$ would result in 
\begin{equation}
    \tilde{H}_{ab}^{\mathrm{MMSE}}(k) = \tilde{H}_{ab}^{\mathrm{min}}(k)\left(\tilde{H}_{ab}^{\mathrm{ap}}(k)\right)^2.
     \label{equ:H_MMSE_dec}
\end{equation}

\section{Simulation Results}\label{sec:simulation}
\begin{table} [t]
\begin{center}
\caption{Simulation parameters} 
\label{tab:sim-parm}
\begin{tabular}{l|l}
\hline Parameters & Specifications \\
\hline \hline FFT Size $N$ & 256 \\
\hline Pilot Rate & $1 / 4$ \\
\hline Guard Interval (CP) & 64 \\
\hline Signal Constellation & $ \mathrm{QPSK}$ \\   
\hline Channel Model & IEEE 802.11 channel model PDP \cite{mimo_ofdm}\\
\hline Max channel taps $L$ & $11$\\
\hline
\end{tabular}
\end{center}
\end{table} 

\begin{figure}[t]
	\centering
	\includegraphics[scale=0.50]{ 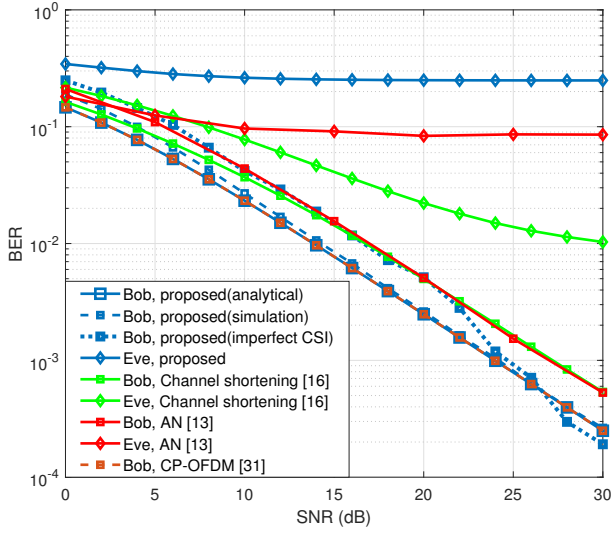}
    	 \footnotesize\caption{ BER performance of the proposed data security algorithm vs channel shortening \cite{8292335}, AN \cite{6516879} and conventional CP-OFDM \cite{mimo_ofdm}.}
	\label{fig:BER_Perfect_CSI}
\end{figure}
In this section, we demonstrate the performance of channel decomposition-based \ac{PLS} algorithms proposed for \ac{OFDM} systems.
To do so, the \ac{BER}-based secrecy gap and \ac{NMSE}-based secrecy gap metrics are used to evaluate the security of data and pilots, respectively. The \ac{BER}-based secrecy gap will quantify the amount of information leakage to the eavesdropper, evaluate the secrecy, and also shows the effect of the proposed algorithm on the reliability with respect to legitimate nodes \cite{8093591}. On the other hand, \ac{NMSE}-based secrecy gap shows the difference between the quality of estimated channel at the legitimate node and illegitimate node. Moreover, the effect of the proposed algorithm on the \ac{PAPR} is also presented along with the comparison to the conventional algorithms. 
The simulated \ac{OFDM} system parameters are described in Table \ref{tab:sim-parm}.

Fig. \ref{fig:BER_Perfect_CSI} shows the \ac{BER} performance versus \ac{SNR} of the proposed data security algorithm along with the comparison to the conventional algorithms such as channel shortening \cite{8292335}, AN \cite{6516879} and conventional CP-OFDM \cite{mimo_ofdm}. First, it is observed that the derived analytical results match well with the simulations. Also, note that under perfect channel estimation the proposed scheme performs exactly as the CP-OFDM while Eve suffers from high error rates. This implies that the proposed data scheme does not degrade the performance of the legitimate user.
Moreover, our novel design exhibits a significant \ac{BER} gap performance compared to AN and channel shortening by ensuring the lowest \ac{BER} values at Bob and the highest error rate at Eve. This result emphasizes the effectiveness of the data security algorithm in maintaining high secrecy levels while preserving the performance of the legitimate user.
Fig. \ref{fig:BER_Perfect_CSI} also shows the BER performance of Bob under imperfect channel estimation. It is seen that at low SNR values, Bob has almost the same performance as AN and channel shortening while at high SNR values it performed slightly better.

\begin{figure}[t]
	\centering
	\includegraphics[scale=0.50]{ 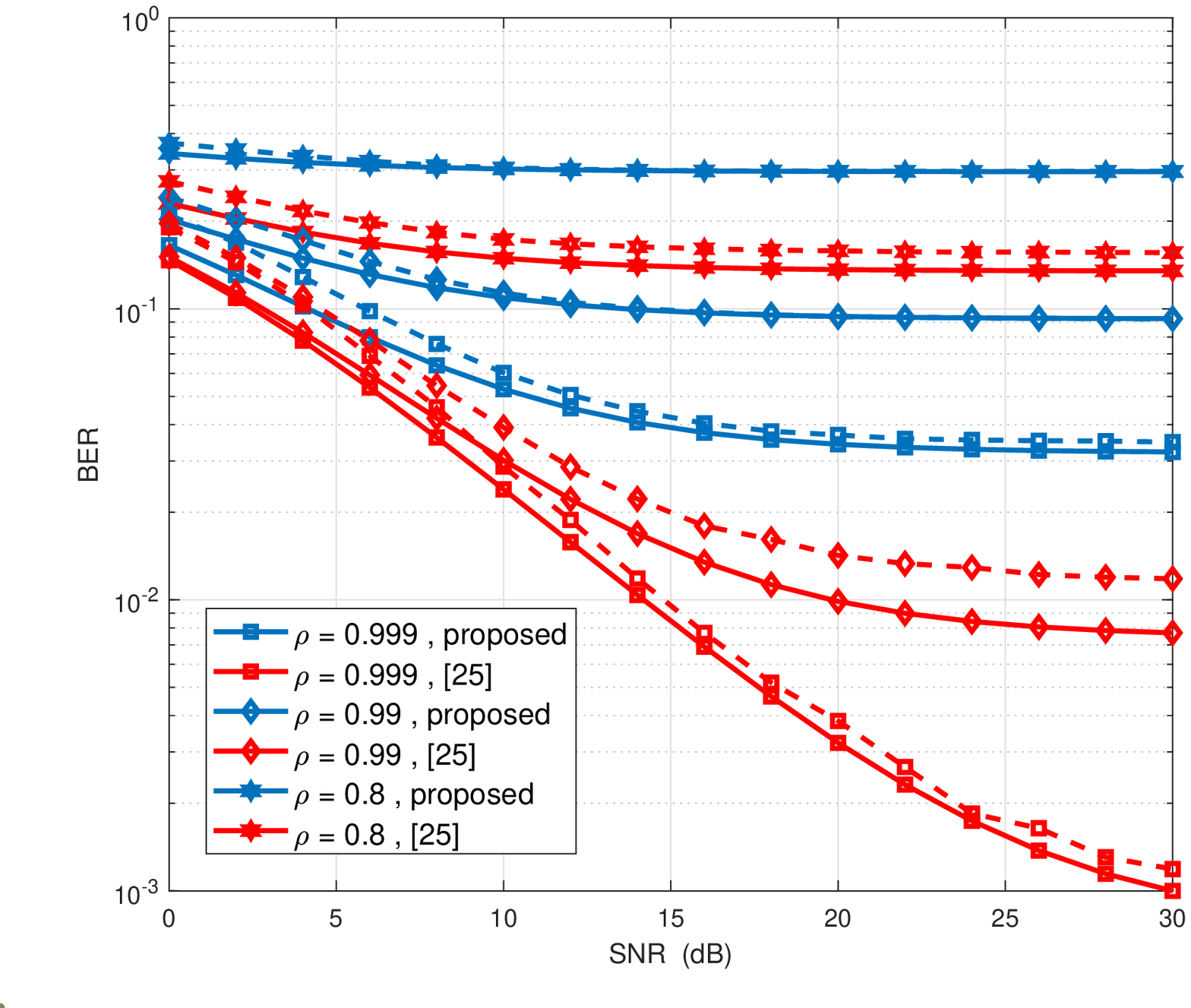}
    	\footnotesize\caption{BER performance of the proposed algorithm (blue color) vs LS \cite{9095399} (red color) under correlated eavesdropping channels. The solid and dashed lines stand for the analytical and simulation results, respectively.}
	\label{fig:Corr_999_99_8}
\end{figure}
\begin{figure}[t]
	\centering
	\includegraphics[scale=0.49]{ 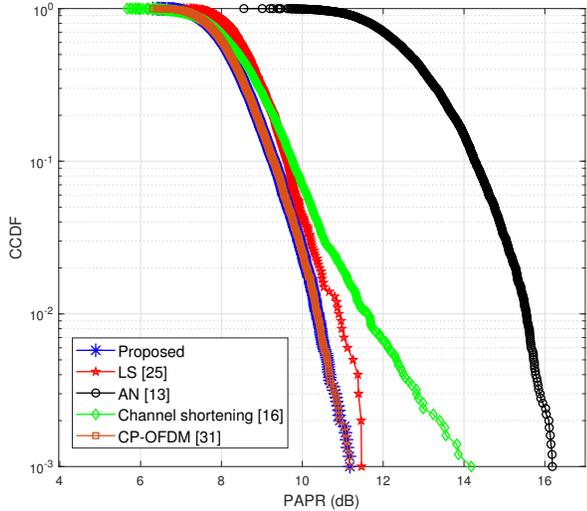}
    	\footnotesize\caption{The CCDF of the PAPR for the proposed algorithm compared to channel shortening \cite{8292335}, AN \cite{6516879}, LS \cite{9095399}, and conventional CP-OFDM \cite{mimo_ofdm}.}
	\label{fig:papr}
\end{figure}
Fig. \ref{fig:Corr_999_99_8} illustrates the analytical as well as the simulated results for \ac{BER} performance of the proposed algorithm and \ac{LS} based algorithm in \cite{9095399} for the case when Eve's channel is correlated with Bob's channel. The analytical results agree well with the simulations confirming the model developed in Subsection \ref{Subsec:Data security analysis}. 
The \ac{BER} performance is evaluated for the correlation values of $\rho = \{0.999, 0.99, 0.8\}$. It is observed that the performance of the Eve improves as the correlation values increase. However, the \ac{BER} performance difference of Eve between the proposed algorithm and LS is large for the same correlation value. For instance, when $\rho =0.999$ the error floor of the proposed scheme settles at $0.03$ while if LS is used the error falls below $0.001$. This is due to the fact that the proposed algorithm is using one component of the channel instead of the overall one, which provides resilience against eavesdroppers near the legitimate nodes as explained in Subsection \ref{Subsec:Data security analysis}.

Thanks to the unit power property of the all-pass channel as explained in Subsection \ref{Subsec:Channel decomp}, the proposed precoding would not cause any power issues since it only changes the phase of each OFDM subcarrier. 
Fig. \ref{fig:papr} depicts \ac{PAPR} performance of the proposed algorithm compared to CP-\ac{OFDM}, AN, LS, and channel shortening methods. It is observed that the \ac{PAPR} performance of the proposed algorithm is similar to that of conventional \ac{OFDM} while the application of other security schemes causes an increase in the \ac{PAPR}. This result makes the usage of the proposed precoding technique independent of any power constraint.

\begin{figure}[h]
	\centering
	\includegraphics[scale=0.49]{ 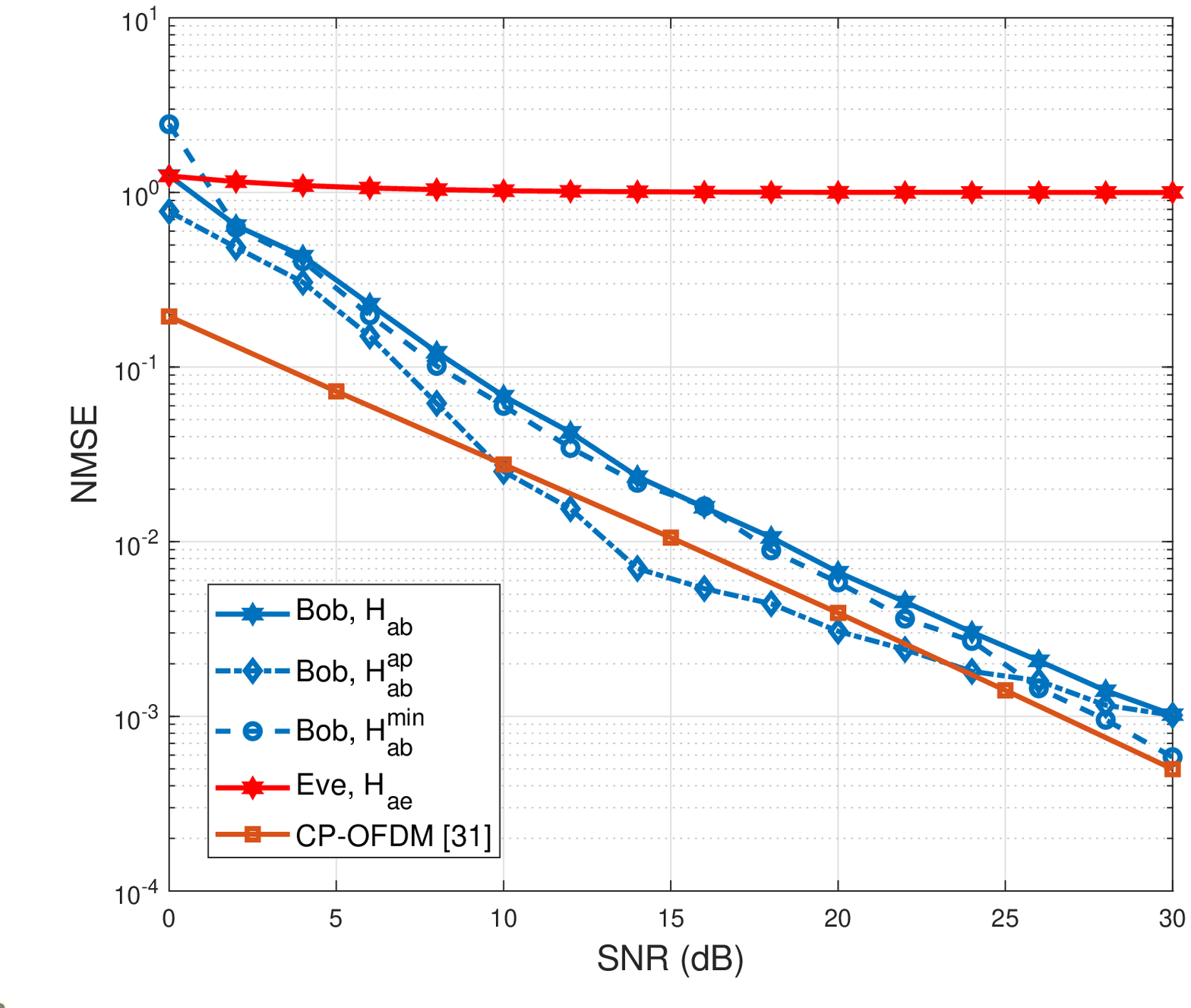}
    	\footnotesize \caption{The channel estimation's NMSE performance vs SNR at Bob and Eve.}
	\label{fig:NMSE}
\end{figure}

Fig. \ref{fig:NMSE} shows the \ac{NMSE} versus \ac{SNR} performance of the estimated channel at the legitimate node and at Eve for the proposed algorithm and CP-OFDM. It is observed that there is a significant estimation error gap between the legitimate and the attacker at the cost of some degradation in the estimation quality at very low SNR values. This ensures that Eve will neither be able to estimate its channel nor acquire the sensing information from the surrounding environment. Moreover, it is also observed that 
the estimation error slightly increases when estimating the minimum-phase or the all-pass channels compared to the effective channel, which justifies the BER performance degradation in Fig. \ref{fig:BER_Perfect_CSI}.

\section{Conclusion and Future Work}\label{sec:conclusion}
In this work, we proposed novel security algorithms for providing data and pilot security. Unlike conventional security schemes which use the full channel, the proposed algorithms decompose the channel into its minimum-phase and all-pass components and exploit only the all-pass part. The latter provides enough randomness to secure the communication without causing any power issues such as high \ac{PAPR} value at the transmitter due to its unit amplitude property. Particularly, the all-pass component and its conjugate are used to secure the pilots and the data, respectively.
For data security, we have considered two scenarios of correlated and uncorrelated eavesdropping channels and evaluated the \ac{BER} gap. Our results reveal that using one component of the channel provides better security than using the total effective channel in both scenarios. Moreover, the results also ensure that the proposed algorithm provides effective security with minimal degradation in the legitimate user’s performance compared to conventional algorithms. For pilot security, we considered the NMSE gap of the estimated channels. The results show that the proposed algorithm is capable of providing significant degradation in the estimated channel quality at the eavesdropper. This will ensure not only securing the CSI but also securing radio environment mapping information. 
Additionally, the proposed algorithm can enable flexibility in terms of providing security to data, pilot, or both depending upon the application requirements. As future work, the proposed algorithm will be investigated with multiple-input multiple-output systems.

\section{Acknowledgment}
The work of H. Arslan was supported by the Scientific and Technological Research Council of Turkey (TUBITAK) under Grant 120C142 and the work of Haji M. Furqan was supported by the HISAR Lab at TUBITAK BILGEM, Gebze, Turkey.  
\begin{appendices}
\section{Proof of \eqref{equ:rho_ae}}\label{App:rho}
Let the correlation function of the legitimate channel gain with the eavesdropping channel be given as
\begin{equation} \label{euq:correlation}
\begin{aligned}
    \rho_{ae} &=\frac{\operatorname{Cov}({H}_{ab}^*,H_{ae})}{\sqrt{\operatorname{var}({H}_{ab}^*)\operatorname{var}(H_{ae})}}\\    
    & = \frac{\operatorname{E}[{H}_{ab}H_{ae}^*]-\operatorname{E}[{H}_{ab}]\operatorname{E}[{H}_{ae}^*]}{\sigma_a\sigma_e}.  
\end{aligned}
\end{equation}
We know that $\operatorname{E}[{H}_{ab}]=\operatorname{E}[{H}_{ae}]=0$. Additionally $H_{ae}$ is given by \eqref{equ:corr_chan}, then
\begin{equation}
\begin{aligned}
    \rho_{ae}& = \frac{\operatorname{E}[{H}_{ab}^*\cdot(\rho_{ae}H_{ab}+\sqrt{1-\rho_{ae}^2}H_{e})]}{\sigma_a\sigma_e}\\  
    & = \frac{\rho_{ae}\operatorname{E}[|{H}_{ab}|^2]+\sqrt{1-\rho_{ae}^2}\left(\operatorname{E}[{H}_{ab}^*]\operatorname{E}[H_{e}]\right)}{\sigma_a\sigma_e}\\ 
    & = \frac{\rho_{ae}\operatorname{E}[|{H}_{ab}|^2]}{\sigma_a\sigma_e}.
\end{aligned}
\end{equation}
Note that $\sigma_a^2= \operatorname{var}(H_{ab})= \operatorname{E}[|{H}_{ab}|^2]-\left(\operatorname{E}[{H}_{ab}]\right)^2 =  \operatorname{E}[|{H}_{ab}|^2]$, and
$\sigma_e^2=\operatorname{var}(\rho_{ae}H_{ab}+\sqrt{1-\rho_{ae}^2}H_{e})=\rho_{ae}^2\sigma_a^2+(1-\rho_{ae}^2)\sigma_a^2=\sigma_a^2$. Therefore, we find
\begin{equation}
\begin{aligned}
    \rho_{ae} = \frac{\rho_{ae}\sigma_a^2}{\sigma_a\sigma_a}=\rho_{ae}.
\end{aligned}
\end{equation}
Thus, we prove that in the channel model given by \eqref{equ:corr_chan}, $\rho_{ae}$ corresponds to the channel correlation between Bob and Eve.

\section{Proof of \eqref{equ:gamma}}\label{App:gamma}

To identify the expression of $\Gamma$ and instead of using complex probability distributions analysis, we exploit the four constraints given in \eqref{equ:const}. For instance, from $C1$, we know that $\Gamma$ is proportional to $\sqrt{1-\rho_{ae}^2}$. And if a linear relationship is assumed, we find 

\begin{equation}
\Gamma = \alpha+\beta \sqrt{1-\rho_{ae}^2},
\end{equation}
where $\alpha$ and $\beta$ are the linear model's parameters to be defined, and by solving the equation in $(C3)$, i.e., $\rho_{ae}(\Gamma=1)=1$, we find the value of the slope as $\alpha=1$.

Also, by using $(C2)$ and $(C4)$ we find the following
\begin{equation}
0\leq\sqrt{1-\rho_{ae}^2}\leq\beta.
\end{equation}
And since $0\leq\sqrt{1-\rho_{ae}^2}\leq1$, we conclude that $\beta=1$, therefore
\begin{equation}
   \Gamma = 1+\sqrt{1-\rho_{ae}^2}.
\end{equation}
This model for $\Gamma$ agrees very well with the simulation results as shown in Fig. \ref{fig:Correl}.

\section{Proof of \eqref{equ:corr_min}}\label{App:corr_min}

The correlation function of the legitimate minimum-phase channel component gain with the eavesdropping channel be given as
\begin{equation}
\begin{aligned}
    &\rho^{\mathrm{min}}_{ae} =\frac{\operatorname{Cov}({H}^{\mathrm{min}*}_{ab},H^{\mathrm{min}}_{ae})}{\sqrt{\operatorname{var}({H}^{\mathrm{min}*}_{ab})\operatorname{var}(H^{\mathrm{min}}_{ae}})}\\    
    & = \frac{\operatorname{E}[{H}^{\mathrm{min}*}_{ab}\cdot H_{ae}^{\mathrm{min}}]-\operatorname{E}[{H}^{\mathrm{min}*}_{ab}]\operatorname{E}[{H}^{\mathrm{min}}_{ae}]}{\sqrt{\left(\operatorname{E}[|{H}^{\mathrm{min}}_{ab}|^2]-\left(\operatorname{E}[{H}^{\mathrm{min}}_{ab}]\right)^2\right)\left(\operatorname{E}[|{H}^{\mathrm{min}}_{ae}|^2]-\left(\operatorname{E}[{H}^{\mathrm{min}}_{ae}]\right)^2\right)}}.  
\end{aligned}
\end{equation}
Note that $\operatorname{E}[{H}^{\mathrm{min}}_{ab}]=\operatorname{E}[{H}^{\mathrm{min}}_{ae}]=0$, $\operatorname{E}[|{H}^{\mathrm{min}}_{ae}|^2]=\operatorname{E}[|{H}_{ae}|^2]=\sigma_e^2$, and $\operatorname{E}[|{H}^{\mathrm{min}}_{ab}|^2]=\operatorname{E}[|{H}_{ab}|^2]=\sigma_a^2$. Please refer to Appendix \ref{App:mean}.\hfill$\blacksquare$

Thus, we find
\begin{equation}
\begin{aligned}
    &\rho^{\mathrm{min}}_{ae} = \frac{\operatorname{E}[{H}^{\mathrm{min}*}_{ab}H_{ae}^{\mathrm{min}}]}{\sigma_a\sigma_e}\\
    & = \frac{\operatorname{E}\bigg[{H}_{ab}^{\mathrm{min}*}\cdot\left(\frac{\rho_{ae}}{1+\sqrt{1-\rho_{ae}^2}}H^{\mathrm{min}}_{ab}+\sqrt{1-\left(\frac{\rho_{ae}}{\Gamma}\right)^2}H^{\mathrm{min}}_{e}\right)\bigg]}{\sigma_a\sigma_e}\\  
    & = \frac{\frac{\rho_{ae}}{1+\sqrt{1-\rho_{ae}^2}}\operatorname{E}[|{H}^{\mathrm{min}}_{ab}|^2]+\sqrt{1-\left(\frac{\rho_{ae}}{\Gamma}\right)^2}\left(\operatorname{E}[{H}^{\mathrm{min}*}_{ab}]\operatorname{E}[H^{\mathrm{min}}_{e}]\right)}{\sigma_a\sigma_e}\\ 
    & = \frac{\frac{\rho_{ae}}{1+\sqrt{1-\rho_{ae}^2}}\sigma_a^2}{\sigma_a\sigma_e}.
\end{aligned}
\end{equation}

Finally, the correlation is found as
\begin{equation}
\rho^{\mathrm{min}}_{ae} =\frac{\rho_{ae}}{1+\sqrt{1-\rho_{ae}^2}}<\rho_{ae}.
\end{equation}

\section{Proof of $\operatorname{E}[{H}^{\mathrm{min}}_{ab}]=\operatorname{E}[{H}^{\mathrm{min}}_{ae}]=0$.} \label{App:mean}

As explained in Subsection \ref{Subsec:Channel decomp}, any FIR causal channel $H_{ab}$ can be decomposed as follows: $H_{ab} = H_{ab}^{\mathrm{min}}\cdot H^{\mathrm{ap}}_{ab}$, where $H^{\mathrm{ap}}_{ab}=e^{jU}$ and $U\sim \mathcal{U}(0,2\pi)$. Thus, $H^{\mathrm{ap}}_{ab}$ has the following \ac{PDF}: $f_X(x)=\frac{1}{j2\pi x}$.

With reference to the main problem, the expectation of $H_{ab}$ is given as
\begin{equation}
\operatorname{E}[{H}_{ab}] = \operatorname{E}[{H}^{\mathrm{min}*}_{ab}\cdot{H}^{\mathrm{ap}}_{ab}] = 0.
\end{equation}

Since the minimum-phase and all-pass components are independent, we find
\begin{equation}
\operatorname{E}[{H}_{ab}] = \operatorname{E}[{H}^{\mathrm{min}}_{ab}]\operatorname{E}[{H}^{\mathrm{ap}}_{ab}] = 0.
\label{equ:proof_exp}
\end{equation}

As shown in \eqref{equ:proof_exp}, whether $\operatorname{E}[{H}^{\mathrm{min}}_{ab}]=0$ or $\operatorname{E}[{H}^{\mathrm{ap}}_{ab}]=0$. However, $\operatorname{E}[{H}^{\mathrm{ap}}_{ab}]=\int_{-\infty}^{+\infty}\frac{x}{j2\pi x}dx \neq 0 $. Therefore
\begin{equation}
\operatorname{E}[{H}^{\mathrm{min}}_{ab}]= 0.
\end{equation}

\end{appendices}


\end{document}